\documentclass[aps,pre,twocolumn]{revtex4-1}
\usepackage[dvips]{graphicx}
\usepackage{amsmath}
\usepackage{siunitx}
\usepackage{float}
\usepackage{verbatim}
\usepackage{color}
\newcommand{\br}{{\bf r}}

\newcommand{\la}{\left\langle}
\newcommand{\ra}{\right\rangle}

\begin{document}

\title{Osmotic Pressure and Swelling Behavior of Ionic Microcapsules}

\author{Mohammed O. Alziyadi}
\altaffiliation[Current address: ]{Department of Mathematics, College of Science and Humanities-Shaqra, 
Shaqra University, Riyadh, Saudi Arabia}
\affiliation{Department of Physics, North Dakota State University,
Fargo, ND 58108-6050, USA}
\author{Alan R. Denton}
\email[]{alan.denton@ndsu.edu}
\affiliation{Department of Physics, North Dakota State University,
Fargo, ND 58108-6050, USA}
\begin{abstract}
Ionic microcapsules are hollow shells of hydrogel, typically 10-1000 nm in radius, 
composed of cross-linked polymer networks that become charged and swollen in a good solvent. 
The ability of microcapsules to swell/deswell in response to changes in external stimuli 
(e.g., temperature, pH, ionic strength) suits them to applications, such as
drug delivery, biosensing, and catalysis. Equilibrium swelling behavior of ionic 
microcapsules is determined by a balance of electrostatic and elastic forces. 
The electrostatic component of the osmotic pressure of a microcapsule -- 
the difference in pressure between the inside and outside of the particle -- plays a 
vital role in determining swelling behavior. Within the spherical cell model, 
we derive exact expressions for the radial pressure profile and for the electrostatic 
and gel components of the osmotic pressure of a microcapsule, which we compute via 
Poisson-Boltzmann theory and molecular dynamics simulation. For the gel component, 
we use the Flory-Rehner theory of polymer networks. By combining the electrostatic 
and gel components of the osmotic pressure, we compute the equilibrium size of 
ionic microcapsules as a function of particle concentration, shell thickness, and valence.
We predict concentration-driven deswelling at relatively low concentrations 
at which steric interactions between particles are weak, and demonstrate that 
this response can be attributed to crowding-induced redistribution of counterions.
Our approach may help to guide the design and applications of smart, 
stimuli-responsive colloidal particles.
\end{abstract}


\maketitle

\section{Introduction}

Microcapsules are soft colloidal particles, typically $10-1000$ nm in radius, 
comprising solvent-permeable shells made of cross-linked polymer (hydrogel)
networks~\cite{Karg-Richtering-langmuir2019,Scotti-hollow-microgels-jcp2018,
Brugnoni-Scotti-nanocapsules2020,Wypysek-Scotti2019} 
or polyelectrolyte multilayers~\cite{Koker_2011_ADDR,DeKoker_2012_CSR,Such_2011_CSR,
Weijun_2012_CSR}. When dispersed in a good, polar solvent, such as water, 
ionic gels become charged by dissociating counterions from the polymer backbones 
and swell by absorbing solvent into the polymer network~\cite{Jia-Muthukumar2021}.
Microcapsules have the additional freedom to swell by absorbing solvent into their cavities.

The degree of swelling 
depends on externally controllable parameters, such as temperature, pH, ionic strength, 
particle concentration, and solvent quality~\cite{ToyoichiTanaka,TongWeijunandDong}. 
For example, variations in temperature can drive a volume phase transition, in which a hydrogel
suddenly collapses upon increasing temperature~\cite{HydrogelBook2012,MicrogelBook2011},
while changes in pH can affect the degree of ionization and thereby the strength of 
electrostatic interactions~\cite{Andelman-Podgornik2021,schneider2017}.
Due to their sensitive and tunable response to changes in environmental conditions, 
microcapsules are of fundamental interest for understanding colloidal self-assembly,
including the glass transition. Moreover, the ability of microcapsules to 
encapsulate and transport fluorescent dyes and drug molecules enables applications,
e.g., to drug delivery~\cite{hamidi2008hydrogel,DeGeest_2009_Softmatter,Schmidt_2011_AFM,
tavakoli2017hydrogel} and optical biosensing~\cite{Kreft_2007_JMC,Borisov_2010_AFR,
McShane_2010_JMC,deMercato_2011_Small,Kazakova_2011_PCCP,RiveraGil_2012_Small} 
for monitoring ion concentrations.

In suspensions of ionic microcapsules (also known as hollow microgels), mobile counterions 
(and coions, if salt is present) redistribute themselves in the solution until local 
Donnan equilibria are established between the cavity, shell, and exterior regions of 
the microcapsules. Differences in concentration of counterions across the interfaces 
separating the cavity from the shell and the shell from the exterior contribute to the 
osmotic pressure of a microcapsule, which in turn affects swelling behavior. In equilibrium,
swelling of these permeable, compressible particles is governed by a delicate balance 
between electrostatic and elastic components of the osmotic pressure. 

The electrostatic component of the osmotic pressure of a microcapsule depends on the 
distributions of free counterions and coions inside, within, and outside the shell
and of immobile ions fixed on the polymer network. The elastic component depends on 
mechanical properties of the polymeric shell and polymer-solvent interactions. 
While previous modeling studies of microcapsules and shells have focused 
on the electrostatic contribution to the osmotic pressure~\cite{Vinogradova_2006_AnnuRevMaterRes,
Stukan_2006_PRE,Vinogradova_2008_JCP,Vinogradova2020,colla-levin2020},
the present study analyzes the interplay between the two contributions, specifically 
for hydrogel microcapsules.

In our previous work on ionic microcapsules, we implemented Poisson-Boltzmann theory 
in the spherical cell model to study the dependence of microcapsule osmotic pressure on 
ionic strength~\cite{Wypysek-Scotti2019} and the dependence of ion density distributions 
and local pH on microcapsule valence, concentration, and permeability~\cite{tang-denton2014,
tang-denton2015}. In related work on ionic microgels, we analyzed the relationship between 
the pressure tensor, osmotic pressure, and equilibrium swelling~\cite{Denton-Alziyadi2019,
Denton-Tang-jcp2016}.
The purpose of the present paper is to further investigate the dependence of osmotic pressure 
and swelling of ionic hydrogel microcapsules on particle concentration, 
architecture, and valence.

The remainder of the paper is organized as follows. Section~\ref{models} first defines
the primitive and cell models of permeable ionic microcapsules. In Sec.~\ref{theory}, 
we then analyze the pressure tensor of ionic hydrogel microcapsules in the cell model 
and derive some exact statistical mechanical relations for the electrostatic component 
of the osmotic pressure of a single particle. In Sec.~\ref{methods}, we discuss two 
computational methods for computing the electrostatic component of the osmotic pressure 
of an ionic microcapsule: Poisson-Boltzmann theory and molecular dynamics simulation.
In Sec.~\ref{results}, after validating our methods by demonstrating close agreement 
between theory and simulation, we calculate equilibrium swelling ratios of ionic 
microcapsules and investigate the dependence on various system parameters.
Finally, in Sec.~\ref{conclusions}, we summarize and conclude with an outlook for future work.

\section{Models}\label{models}

\subsection{Ionic Microcapsule Suspension}\label{suspension}

The system of interest consists of a suspension of $N_m$ ionic hydrogel microcapsules 
dispersed in a solvent of volume $V$ at temperature $T$. For computational efficiency, 
and since we are concerned here only with equilibrium properties,
we adopt the primitive model of macroion solutions, which treats the solvent implicitly 
as a dielectric continuum of uniform dielectric constant $\epsilon$ and considers explicitly 
only the mobile microions. Each microcapsule is a hollow microgel -- a spherical shell 
of hydrogel of dry inner and outer radii $a_0$ and $b_0$, respectively, and swollen inner 
and outer radii $a$ and $b$. The microcapsule number density, $n_m=N_m/V$, is related to 
the volume fraction of swollen particles, $\phi=(4\pi/3)n_m b^3$, and of dry (collapsed) 
particles, $\phi_0=(4\pi/3)n_m b_0^3$. 

The hydrogel is a polymer network comprising monomers bonded to one another in linear chains,
which are cross-linked into subchains. When ionized, the hydrogel carries a fixed charge 
$-Ze$ ($Z$ being the valence and $e$ the electron charge), originating from dissociation 
of counterions from some of the monomers or of a chemical species in the network 
(e.g., initiator in the synthesis reaction). 
Although we assume negative fixed charge, in line with commonly studied 
hydrogels, e.g., poly(N-isopropylacrylamide) (PNIPAM), our results apply also to positive 
fixed charge upon changing the sign of $Z$.
For simplicity, we assume here that the average distribution of fixed charge 
is uniform throughout the shell ($a\le r\le b$) -- a reasonable assumption considering 
thermal motion of flexible, loosely cross-linked polymer chains. With this assumption, 
the average number density of fixed charge is given by
\begin{equation}
n_f(r)=\left\{ \begin{array}
{l@{\quad}l}
0, 
& r<a \\[2ex]
\frac{\displaystyle 3Z}{\displaystyle 4\pi(b^3-a^3)},
& a\le r\le b \\[2ex]
0,
& r>b.
\end{array} \right.
\label{nfr}
\end{equation}

Electroneutrality of the suspension dictates the number of dissociated counterions 
of valence $z$ to be $N_mZ/z$. Salt in solution adds to the total numbers $N_{\pm}$ 
of mobile microions (counterions/coions), which are free to diffuse (along with 
solvent molecules) through the permeable shell of the microcapsule and into the hollow cavity.
We model the microions as monovalent ($z=\pm 1$) point charges, thus 
neglecting steric interactions and correlations between their cores. 
This model is reasonable for salt-free aqueous suspensions -- our focus here -- 
in which electrostatic repulsion ensures that like-charged counterions rarely approach 
so closely that their cores interact. For salty suspensions, however, attraction between
oppositely charged counterions and coions allows for strong core-core interactions.
In the latter case, incorporating nonzero microion size into the model prevents 
microions from approaching arbitrarily closely and accounts for correlations associated
with excluded volume~\cite{colla-levin2020,Ben_Yaakov_2009}. The point charge model 
further neglects steric interactions between microions and monomers making up the
polymer network. This approximation is reasonable for hydrogels with a relatively
low monomer volume fraction in their swollen state, and is consistent with the 
Flory-Rehner mean-field theory of swollen polymer networks (Sec.~\ref{osmotic-pressure}),
but is less justified for dense nanoshells (e.g., vesicles, biological membranes,
and viral capsids)~\cite{colla-levin2020}.

\begin{figure}[t!]
\includegraphics[width=0.9\columnwidth,angle=0]{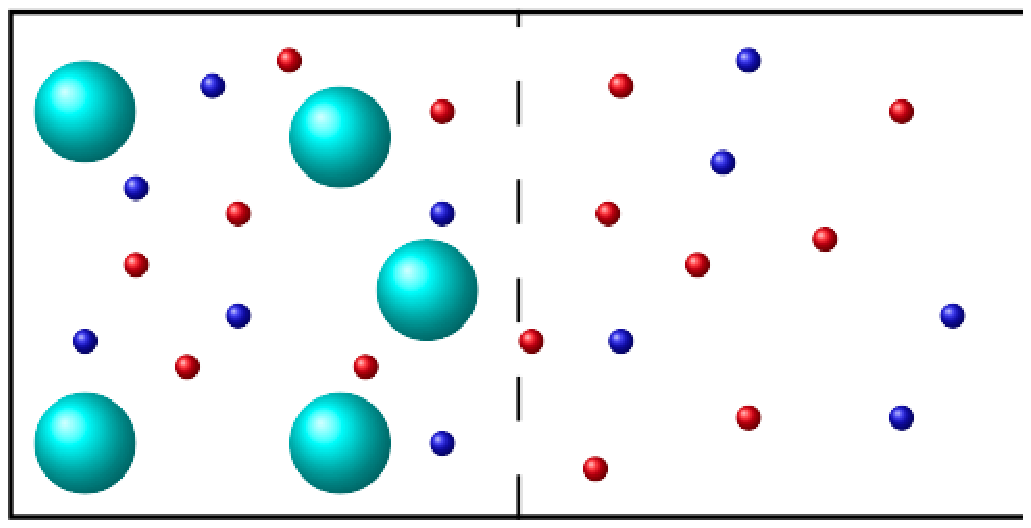}
\vspace*{-0.2cm}
\caption{Primitive model of spherical ionic microcapsules (large cyan spheres) and point 
counterions/coions (small red/blue spheres) dispersed in a solvent, modeled as a 
dielectric continuum. In Donnan equilibrium, microions in the suspension (left)
can exchange with an electrolyte reservoir (right) across a semipermeable membrane
(dashed black line).
}\label{figsus}
\end{figure}

In a closed suspension, all particles are confined to the same volume. In Donnan equilibrium,
the microions can diffuse through a semipermeable membrane separating the suspension from
an electrolyte (salt) reservoir (Fig.~\ref{figsus}). Chemical equilibrium with respect to 
exchange of microions then determines the salt concentration of the suspension. Of the total 
number of microions in the suspension, $N=N_++N_-$, the number of counterions is 
$N_+=ZN_m+N_-$ and the number of coions is $N_-$.

In equilibrium, the microions distribute themselves to equalize their chemical potential.
The resulting nonuniform distribution of macroions and microions between the suspension 
and reservoir generates an osmotic pressure in the suspension -- defined as the difference 
in pressure between the suspension and the reservoir. Microions can similarly exchange 
between the interior and exterior regions of a single microcapsule. The periphery of the 
hydrogel thus acts itself like a semipermeable membrane, allowing mobile microions to 
penetrate, but retaining the fixed charge inside. The resulting nonuniform distribution 
of microions between the inside and outside of a microscapsule establishes a {\it local} 
Donnan equilibrium and a corresponding single-particle osmotic pressure, defined as the 
difference in average pressure between the inside and outside of the gel. 
An important distinction is that, whereas a semipermeable membrane bounding a suspension
of microcapsules may sustain a nonzero osmotic pressure in equilibrium, a single compressible 
microcapsule cannot. As discussed further below in Sec.~\ref{theory}, the microcapsule 
gel swells to an equilibrium size at which the single-particle osmotic pressure vanishes. 

\begin{figure}[t!]
\includegraphics[width=0.9\columnwidth,angle=0]{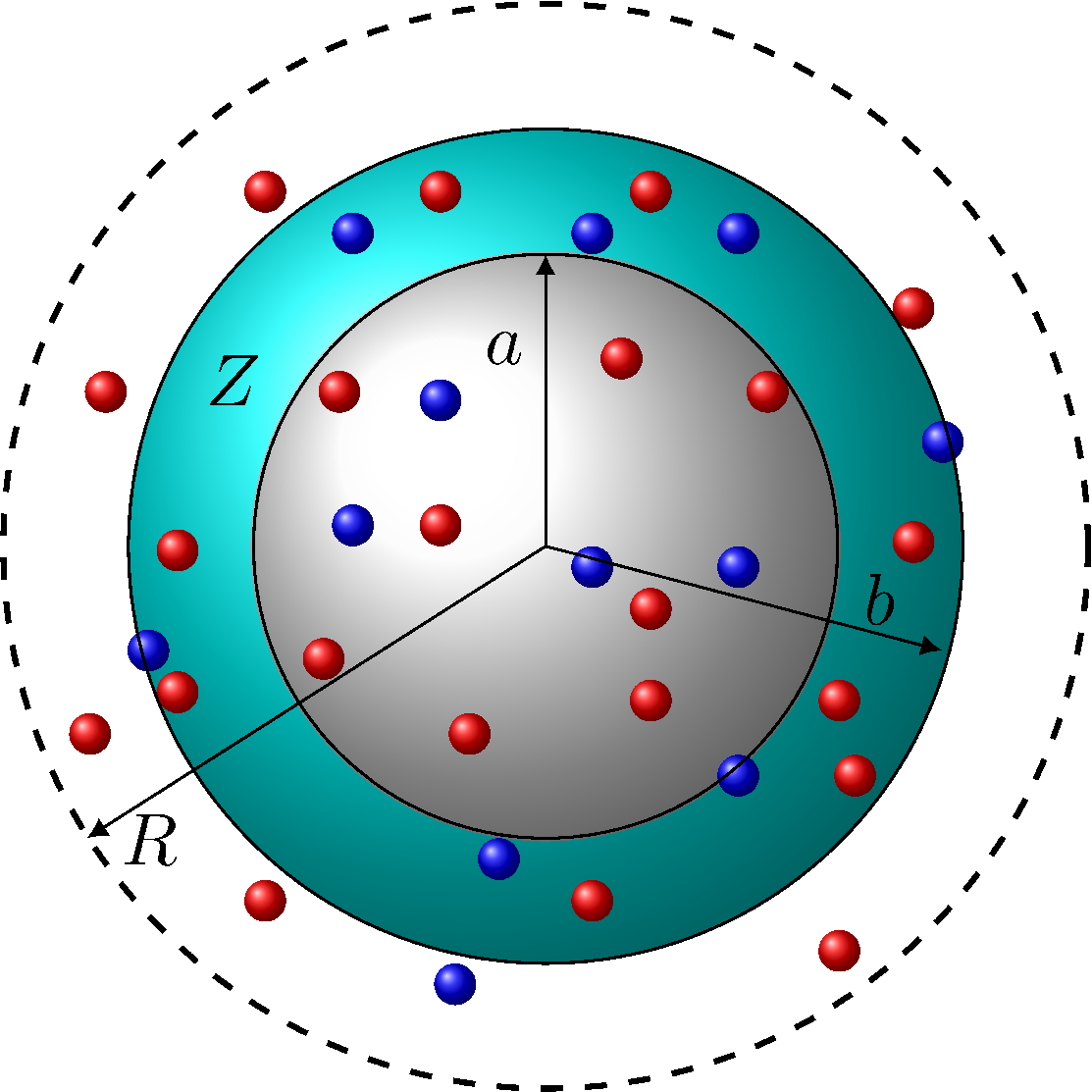}
\vspace*{-0.2cm}
\caption{Schematic representation of the cell model. A charged microcapsule 
(cyan shell shown in cross-section), of inner and outer swollen radii $a$ and $b$, 
is placed at the center of a spherical cell. The cell radius $R$ is fixed by the 
volume fraction $\phi=(b/R)^3$ of microcapsules in the suspension. 
In local Donnan equilibrium, mobile counterions and coions (small red and blue spheres)
can exchange between the solution, shell, and solvent-filled cavity (grey sphere).}
\label{figmodel}                                                                              
\end{figure}

\subsection{Spherical Cell Model}

To model ionic microcapsules in a suspension, we adopt a version of the cell model.
The cell model reduces the complexity of a bulk suspension with many interacting macroions 
to the relative simplicity of a single macroion centered in a cell of like geometry
(Fig.~\ref{figmodel}), at the expense of neglecting correlations between macroions. 
For spherical microcapsules, the cell, which must have the same symmetry, has a 
radius $R$ determined by the volume fraction $\phi_0=(b_0/R)^3$ of dry (collapsed) 
microcapsules in the suspension. In their swollen state, the microcapsules occupy 
a volume fraction $\phi=(b/R)^3=\phi_0\alpha^3$, 
thus defining the swelling ratio $\alpha\equiv b/b_0$ as the ratio of the outer swollen 
and dry radii. Although the inner and outer swollen radii may, in principle,
vary independently, we assume, for simplicity, uniform swelling of the microcapsule, 
such that $\alpha=b/b_0=a/a_0$.

\section{Theory}\label{theory} 

To predict the equilibrium swollen size of ionic microcapsules in a bulk suspension, 
we begin by analyzing the pressure tensor ${\bf P}$ in the spherical cell model.
We then discuss an equivalent statistical mechanical approach. Preliminary derivations 
of these complementary theoretical approaches were presented 
in ref.~\cite{Wypysek-Scotti2019}. 

\subsection{Osmotic Pressure in the Cell Model}\label{osmotic-pressure} 

In continuum mechanics~\cite{landau-lifshitz1984}, the pressure tensor 
(negative of the stress tensor) at position ${\bf r}$ in a material 
at temperature $T$ is defined by its matrix elements,
\begin{equation}
P_{ij}({\bf r})\equiv -\left(\frac{\delta{\cal F}}{\delta\eta_{ij}({\bf r})}\right)_T,
\label{P-def}
\end{equation}
which are functional derivatives of the Helmholtz free energy functional 
${\cal F}[\eta({\bf r})]$ with respect to the elements $\eta_{ij}({\bf r})$ 
of the local deformation (strain) tensor $\eta({\bf r})$. In the spherical cell model, 
owing to spherical symmetry, ${\bf P}$ is represented by a diagonal matrix,
\begin{equation}
\left({\bf P}\right)=
\left(
\begin{array}{ccc} 
P(r) & 0 & 0 \\
0 & P_{\bot}(r) & 0 \\
0 & 0 & P_{\bot}(r)
\end{array}
\right),
\label{P-matrix}
\end{equation}
whose radial and tangential elements, $P(r)$ and $P_{\bot}(r)$, respectively, 
depend on only the radial distance $r$ from the center of the cell. We focus 
on the radial element, since it determines radial swelling behavior of microcapsules.
For comparison, the pressure of the suspension corresponds to the value of $P(r)$ 
at the cell edge $P(R)$. 

To make progress, it is convenient to decompose $P(r)$ into two physically distinct
contributions -- a gel component $P_g(r)$, associated with elasticity and solvation 
of the polymer network, and an electrostatic component $P_e(r)$, associated with 
electrostatic forces among localized (fixed) ions in the network and mobile microions: 
\begin{equation}
P(r, \alpha)=P_g(r, \alpha)+P_e(r, \alpha).
\label{Ptot}
\end{equation}
For once, we make explicit that both components of $P(r)$ depend implicitly on 
the microcapsule swelling ratio $\alpha$.
In thermodynamic equilibrium, when the microcapsules achieve their equilibrium
swollen size, mechanical stability dictates that the divergence of the pressure tensor 
vanishes, $\nabla\cdot{\bf P}=0$, and thus $P(r)$ is continuous throughout 
the cell~\cite{landau-lifshitz1984,trizac-hansen1997,Widom2009,Denton-Alziyadi2019}, 
including at the inner and outer edges of the microcapsule shell 
($r=a$ and $b$ in Fig.~\ref{figmodel}).
It is important to emphasize that for bulk gels the pressure tensor is spatially 
uniform~\cite{Jia-Muthukumar2021}, while for microgels in the cell model, 
${\bf P}$ is spatially nonuniform~\cite{trizac-hansen1997,Widom2009,Denton-Alziyadi2019}.
Substituting Eq.~(\ref{P-matrix}) into the general expression for the divergence 
of a tensor in spherical polar coordinates~\cite{RHB,sochi2016} yields
\begin{equation}
\nabla\cdot{\bf P}(r)=\frac{\partial P(r)}{\partial r}
+\frac{2}{r}[P(r)-P_{\bot}(r)]=0.
\label{divergence-sphere}
\end{equation}

The radial element of the electrostatic component of the pressure tensor 
is formally given by~\cite{landau-lifshitz1984,trizac-hansen1997,Widom2009}
\begin{equation}
\beta P_e(r)=\beta P_{\rm kin}-\frac{|\psi'(r)|^2}{8\pi\lambda_B},
\label{Per}
\end{equation}
comprising the kinetic pressure $P_{\rm kin}$ of the mobile microions 
and the Maxwell pressure (second term) generated by the total electric field, 
$E(r)=-\varphi'(r)$. Here $\psi(r)\equiv\beta e\varphi(r)$ is the reduced (dimensionless) 
form of the electrostatic potential $\varphi(r)$, $\beta\equiv 1/(k_BT)$ 
($k_B$ is the Boltzmann constant), and $\lambda_B=\beta e^2/\epsilon$ 
is the Bjerrum length, defined as the distance at which the Coulomb 
interaction energy between two counterions equals the thermal energy $k_BT$.
As further consequences of spherical symmetry, the electric field has only 
a radial component and both $E(r)$ and $n_{\pm}(r)$ are radially symmetric.
In a mean-field approximation that neglects correlations among microions, 
$\beta P_{\rm kin}=n_+(r)+n_-(r)$, where $n_{\pm}(r)$ are average microion 
number density radial profiles.

In stable mechanical equilibrium, the gel component of the pressure tensor 
has a radial element equal to the local pressure that the gel must exert 
in the radial direction to oppose the local radial electrostatic pressure 
and maintain an immobile distribution of fixed charge. Since the local 
force density is proportional to the electric field (i.e., $-\psi'(r)$) 
times the fixed charge density, integration over the radial coordinate yields
\begin{equation}
\beta P_g(r)=-\int_r^R du\, n_f(u)\psi'(u).
\label{Pgr1}
\end{equation}
Elsewhere, we showed that this relation can be independently derived by minimizing 
a free energy functional within Poisson-Boltzmann theory~\cite{Denton-Alziyadi2019}.
Upon substituting the fixed charge distribution $n_f(r)$ of a swollen microcapsule
from Eq.~(\ref{nfr}), we obtain
\begin{equation}
\beta P_g(r)v_{\rm sh}=Z\left\{ \begin{array}
{l@{\hspace{2mm}}l}
\psi(a)-\psi(b),
& r<a \\[2ex]
\psi(r)-\psi(b),
& a\le r\le b \\[2ex]
0,
& r>b,
\end{array} \right.
\label{Pgr2}
\end{equation}
where $v_{\rm sh}=(4\pi/3)(b^3-a^3)$ is the volume of the shell.

Another useful interpretation of the equilibrium condition of continuity 
of the pressure tensor follows from considering the difference between the 
radial elements of the electrostatic component $P_e(r)$ and $P(r)$ itself:
\begin{equation}
\Delta P_e(r)\equiv P_e(r)-P(r)=-P_g(r),
\label{Delta-Per1}
\end{equation}
from which the equilibrium continuity condition becomes
\begin{equation}
\Delta P_e(r)+P_g(r)=0.
\label{Delta-Per2}
\end{equation}
Averaging over the microcapsule volume, $v=(4\pi/3)b^3$, and defining
\begin{equation}
\pi_e(\alpha)\equiv\frac{4\pi}{v}\int_0^b dr\, r^2 \Delta P_e(r)
\label{pie}
\end{equation}
and
\begin{equation}
\pi_g(\alpha)\equiv\frac{4\pi}{v}\int_0^b dr\, r^2 P_g(r)
\label{pig1}
\end{equation}
as electrostatic and gel components, respectively, of the osmotic pressure $\pi(\alpha)$ 
of a single microcapsule, the equilibrium condition can be expressed as
\begin{equation}
\pi(\alpha_{\rm eq})=\pi_g(\alpha_{\rm eq})+\pi_e(\alpha_{\rm eq})=0,
\label{pitot}
\end{equation}
where $\alpha_{\rm eq}$ denotes the equilibrium value of the swelling ratio.
It is essential here to distinguish $\pi(\alpha)$ from the osmotic pressure 
of the suspension, defined as the change in $P(r)$ between the cell boundary
($r=R$ in Fig.~\ref{figmodel}) and the electrolyte reservoir.
As discussed in our previous work on osmotic pressure and swelling of
ionic microgels~\cite{Denton-Alziyadi2019,Denton-Tang-jcp2016} and
below in Sec.~\ref{stat-mech}, an alternative, but equivalent, expression 
for $\pi_e(\alpha)$ can be derived from a statistical mechanical approach and 
a thermodynamic definition of pressure~\cite{colla-likos2014}.

An independent relation between the gel component of the microcapsule osmotic pressure 
and the swelling ratio can be obtained from theories or simulations of nonionic microgels.
For purposes of illustration, we adopt here the Flory-Rehner theory of polymer 
networks~\cite{flory1953,flory-rehner1943-I,flory-rehner1943-II}. 
Combining mean-field approximations for the osmotic pressure contribution 
associated with mixing entropy and polymer-solvent interactions, 
\begin{equation}
\beta\pi_{\rm mix}(\alpha)v_{\rm sh}=-N_{\rm mon}[\alpha^3\ln(1-\alpha^{-3})+\chi\alpha^{-3}+1],
\label{pi-mixing}
\end{equation}
and for the contribution from elastic network energy, 
\begin{equation}
\beta\pi_{\rm elastic}(\alpha)v_{\rm sh}=-N_{\rm ch}(\alpha^2-1/2),
\label{pi-elastic}
\end{equation}
the Flory-Rehner theory predicts 
\begin{eqnarray}
\beta\pi_g(\alpha)v_{\rm sh}=&-&N_{\rm mon}[\alpha^3\ln(1-\alpha^{-3})+\chi\alpha^{-3}+1]
\nonumber\\[1ex]
&-&N_{\rm ch}(\alpha^2-1/2),
\label{pig2}
\end{eqnarray}
where $N_{\rm mon}$ is the number of monomers, $N_{\rm ch}$ is the number of chains, 
and $\chi$ is the Flory solvency parameter. By equating this expression with
$\pi_g(\alpha)$ from Eqs.~(\ref{Pgr2}) and (\ref{pig1}), we can solve for 
an ionic microcapsule's equilibrium swelling ratio. 

Evidently, practical implementation of the approach outlined above requires 
determining the electrostatic potential in the cell. In Sec.~\ref{PBtheory} below, 
we discuss a computational method for computing $\psi(r)$ based on Poisson-Boltzmann theory.
First, however, we derive an alternative, but equivalent, statistical mechanical 
expression for the electrostatic component of the osmotic pressure of 
a microcapsule in the cell model.

\subsection{Exact Statistical Mechanical Relations}\label{stat-mech} 

In a second theoretical approach to modeling swelling of ionic microcapsules, we adapt 
the statistical mechanical method developed in refs.~\cite{Denton-Alziyadi2019} and
\cite{Denton-Tang-jcp2016}. This approach, which is exact in the cell model, 
determines the average electrostatic contribution to the osmotic pressure by 
considering the change in the electrostatic free energy $F_e$ of a microcapsule 
upon a virtual change of its volume:
\begin{equation}
\pi_e=-\left(\frac{\partial F_e}{\partial v}\right)_{\gamma, T}
=-\frac{1}{4\pi b^2}\left(\frac{\partial F_e}{\partial b}\right)_{\gamma, T},
\label{pie1}
\end{equation} 
where the radius ratio $\gamma\equiv a/b=a_0/b_0$ is held constant when taking the 
derivative, consistent with our assumption of uniform swelling. The latter assumption
implies that the electrostatic contribution to the {\it interior} osmotic pressure 
of a microcapsule $\pi_{{\rm in}, e}$ -- defined as the pressure difference across 
the interface between the inside surface of the hydrogel shell and the hollow cavity 
-- is directly related to the exterior electrostatic osmotic pressure via 
\begin{equation}
\pi_{{\rm in},e}=-\frac{1}{4\pi a^2}\left(\frac{\partial F_e}{\partial a}\right)_{\gamma}
=-\frac{1}{\gamma^3}\pi_e.
\label{pie-in}
\end{equation} 
In previous related work, Colla, Likos, and Levin~\cite{colla-likos2014} invoked a 
similar thermodynamic definition of the electrostatic osmotic pressure [Eq.~(\ref{pie1})],
which they evaluated within Poisson-Boltzmann (PB) theory to analyze the dependence on 
salt concentration of swelling of regular ionic microgels (with no cavity). Relying on 
knowledge of the free energy, however, such an approach is necessarily limited to 
thermodynamic theories (e.g., PB theory) that predict free energy). In contrast, 
our statistical mechanical approach, as shown below, leads to a general expression 
for $\pi_e$ that is independent of PB theory.

When compared with the definition of $\pi_e$ in Eq.~(\ref{pie}), derived 
from the pressure tensor, the thermodynamic definition in Eq.~(\ref{pie1}) 
suggests [via Eq.~(\ref{P-def})] the relation
\begin{equation}
\left(\frac{\partial F_e}{\partial v}\right)_{\gamma, T}
=\frac{1}{v}\int_v dv\,
\left(\frac{\delta{\cal F}_e}{\delta\eta_{rr}({\bf r})}\right)_{\gamma, T},
\label{pie-def}
\end{equation}
where ${\cal F}_e[\eta({\bf r})]$ is the electrostatic contribution to the 
free energy functional, whose equilibrium value is $F_e$, and $\eta_{rr}$ 
is the $rr$ element of the matrix representing the deformation tensor. 
Although we do not prove this conjectured relation here, our numerical results, 
presented below in Sec.~\ref{results}, strongly support its validity
in the cell model.

In a closed suspension, the electrostatic Helmholtz free energy is given by
\begin{equation}
F_e=-k_BT\ln\la\exp(-\beta H_e)\ra,
\label{Fe}
\end{equation}
where angular brackets denote a trace over microion configurations 
in the canonical ensemble and
\begin{equation}
H_e= U_m(b,\gamma)+U_{m\mu}(\{\br\};b,\gamma)+U_{\mu\mu}(\{\br\})
\label{He}
\end{equation}
is the electrostatic component of the Hamiltonian. The latter
decomposes naturally into a microcapsule electrostatic self-energy $U_m(b,\gamma)$ 
and microcapsule-microion and microion-microion electrostatic interaction energies, 
$U_{m\mu}(\{\br\};b,\gamma)$ and $U_{\mu\mu}(\{\br\})$, respectively, which depend on 
the coordinates $\{\br\}\equiv \{{\bf r}_1,\ldots,{\bf r}_N\}$ of all $N$ microions. 
Substituting Eq.~(\ref{He}) into Eqs.~(\ref{pie1}) and (\ref{Fe}),
and noting that $U_{\mu\mu}(\{\br\})$ is independent of $b$, we obtain
\begin{equation}
\pi_e=-\frac{1}{4\pi b^2}\left[\left(\frac{\partial U_m(b,\gamma)}{\partial b}\right)_{\gamma}
+\la\frac{\partial U_{m\mu}(\{\br\};b,\gamma)}{\partial b}\ra_{\gamma}\right].
\label{pie2} 
\end{equation} 
Within our coarse-grained model of a microcapsule, explicit expressions for 
$U_m(b,\gamma)$ and $U_{m\mu}(\{\br\};b,\gamma)$ can be derived from the 
electric field $E_f(r)=-\varphi_f'(r)$ and electrostatic potential $\varphi_f(r)$ 
generated by a spherical shell of uniformly distributed fixed charge:
\begin{equation}
E_f(r)=
\left\{\begin{array}{cc} \vspace{3mm} 
0, & \hspace{2mm} 0<r\le a\\ \vspace{3mm} 
-\frac{\displaystyle Ze(r^3-a^3)}{\displaystyle\epsilon(b^3-a^3)r^2}, & \hspace{2mm} a<r\le b \\ \vspace{3mm}
-\frac{\displaystyle Ze}{\displaystyle\epsilon r^2}, & \hspace{2mm} r>b,\\ 
\end{array} \right.
\label{Efr}
\end{equation}
\begin{equation}
\varphi_f(r)=
\left\{\begin{array}{cc} \vspace{3mm} 
-\frac{\displaystyle 3Ze(b^2-a^2)}{\displaystyle 2\epsilon(b^3-a^3)}, & \hspace{2mm} 0<r\le a\\ \vspace{3mm} 
-\frac{\displaystyle Ze}{\displaystyle \epsilon(b^3-a^3)}
\left(\frac{\displaystyle 3b^2}{\displaystyle 2}-\frac{\displaystyle r^2}{\displaystyle 2}
-\frac{\displaystyle a^3}{\displaystyle r}\right), & \hspace{2mm} a<r\le b\\ \vspace{3mm} 
-\frac{\displaystyle Ze}{\displaystyle\epsilon r}, & \hspace{2mm} r>b.\\ 
\end{array} \right.
\label{phir}
\end{equation}
Substituting Eq.~(\ref{Efr}) into the relation for the energy stored in an electric field,
\begin{equation}
U_m(b,\gamma)=\frac{\epsilon}{2}\int_0^R dr\, r^2|E_f(r)|^2,
\label{Um1}
\end{equation}
yields the self-energy of the fixed charge:
\begin{equation}
\beta U_m(b,\gamma)=\frac{3Z^2\lambda_B}{10b}~\frac{2-5\gamma^3+3\gamma^5}{(1-\gamma^3)^2},
\label{Um2}
\end{equation}
and thus
\begin{equation}
\beta\left(\frac{\partial U_m}{\partial b}\right)_{\gamma}=
-\frac{3Z^2\lambda_B}{10b^2}~\frac{2-5\gamma^3+3\gamma^5}{(1-\gamma^3)^2},
\label{dUmdb}
\end{equation}
which gives [via Eq.~(\ref{pie2})] the self-energy contribution to the 
electrostatic component $\pi_e$ of the osmotic pressure of a microcapsule.
Substituting Eq.~(\ref{phir}) into the relation for the energy of interaction 
between the microcapsule and the microions,
\begin{equation}
U_{m\mu}(\{\br\};a,b)=e\sum_{i=1}^N z_i\varphi_f(r_i),
\label{Ummu1}
\end{equation}
where $z_i=\pm 1$ is the valence of microion $i$ and $r_i$ the radial distance 
of microion $i$ from the cell center, and then taking a derivative with respect to $b$, 
we obtain
\begin{eqnarray}
&&\hspace*{-0.4cm}\beta\left(\frac{\partial}{\partial b}U_{m\mu}\right)_{\gamma}=
\frac{3Z\lambda_B}{2b^2(1-\gamma^3)}\Bigl[\la N_+\ra_{\rm sh}-\la N_-\ra_{\rm sh}
\nonumber\\[1ex]
&-&\frac{1}{b^2}\left(\la r_+^2\ra_{\rm sh}-\la r_-^2\ra_{\rm sh}\right)
+(1-\gamma^2)\left(\la N_+\ra_{\rm cav}-\la N_-\ra_{\rm cav}\right)\Bigr],
\nonumber\\[1ex]
\label{dUmmudb}
\end{eqnarray}
which gives [via Eq.~(\ref{pie2})] the macroion-microion interaction contribution 
to the electrostatic component of the osmotic pressure of a single macroion.
Here, for microion number density profiles $n_{\pm}(r)$, 
\begin{equation}
\la N_{\pm}\ra_{\rm cav}=4\pi\int_0^a dr\, r^2 n_{\pm}(r)
\label{Npmcav}
\end{equation}
and
\begin{equation}
\la N_{\pm}\ra_{\rm sh}=4\pi\int_a^b dr\, r^2 n_{\pm}(r)
\label{Npmsh}
\end{equation}
denote the mean numbers of counterions/coions in the cavity and shell regions, 
respectively, and 
\begin{equation}
\la r_{\pm}^2\ra_{\rm sh}=4\pi\int_a^b dr\, r^4 n_{\pm}(r)
\label{r2pmsh}
\end{equation}
denote mean-square radial distances of microions within the shell.
Finally, combining Eqs.~(\ref{pie2}), (\ref{dUmdb}), and (\ref{dUmmudb}), we obtain
the electrostatic component of the osmotic pressure of a microcapsule:
\begin{eqnarray}
\beta\pi_e v_{\rm sh}
&=&\frac{Z\lambda_B}{2b}\Bigl[\frac{Z}{1-\gamma^3}\left(\frac{2}{5}-\gamma^3
+\frac{3}{5}\gamma^5\right) \nonumber\\[1ex]
&-&\la N_+\ra_{\rm sh}+\la N_-\ra_{\rm sh}
+\frac{1}{b^2}\left(\la r_+^2\ra_{\rm sh}-\la r_-^2\ra_{\rm sh}\right) \nonumber\\[1ex]
&-&(1-\gamma^2)\left(\la N_+\ra_{\rm cav}-\la N_-\ra_{\rm cav}\right)\Bigr].
\label{pie3}
\end{eqnarray}
We emphasize that this expression, which generalizes to microcapsules the expression 
first derived in ref.~\cite{Denton-Tang-jcp2016} for ionic microgels (without a cavity),
is {\it exact} within the combined primitive and cell models.
As it does not require knowledge of the free energy, our approach 
to calculating the electrostatic component of the osmotic pressure
is not tied to PB theory~\cite{colla-likos2014}. In fact, our expression for $\pi_e$ 
[Eq.~(\ref{pie3})] may be evaluated from any statistical mechanical theory or 
molecular simulation method. Moreover, our approach directly connects the mechanical 
pressure tensor to the thermodynamic osmotic pressure. 
For applications in which the nonzero microion sizes are important, 
however, additional contributions to the total osmotic pressure must be included to
account for steric interactions and excluded-volume 
correlations~\cite{colla-levin2020}.

As a consistency check, we note that Eq.~(\ref{pie3}) reduces to the result for 
regular microgels in the limit $a\to 0$.
Furthermore, in the limit of an infinitesimally thin shell ($a\to b$ or $\gamma\to 1$),
i.e., for a fixed charge uniformly distributed over the outer surface, 
Eq.~(\ref{pie3}) reduces to the relatively simple result
\begin{equation}
\beta\pi_e=\frac{\lambda_BZ(Z-2Z_{\rm in})}{8\pi a^4},
\label{pie_shell}
\end{equation}
where 
\begin{equation}
Z_{\rm in}\equiv\la N_+\ra_{\rm cav}-\la N_-\ra_{\rm cav}
\label{Znet}
\end{equation}
is the average net valence of the microcapsule interior ($r<a$). This result, 
following directly from Eq.~(\ref{pie3}) upon setting $\gamma=1-\delta$ and evaluating 
the $\delta\to 0$ limit using L'H\^opital's rule, agrees with the expression for the 
electrostatic contribution to the osmotic stress of surface-charged nanoshells derived 
by Colla, Bakhshandeh, and Levin~\cite{colla-levin2020}. These authors also derived,
via classical density-functional theory, additional contributions from steric interactions 
between microions and the shell walls, which can be significant in the case
of dense nanoshells. Note that, in opposition to the positive self-energy contribution,
the microions make a net negative contribution to the electrostatic component of the 
osmotic pressure, since $\la N_+\ra_{\rm cav}>\la N_-\ra_{\rm cav}$ for negatively 
charged hydrogels, thus promoting deswelling of microcapsules.

Next, we discuss theoretical and simulation methods for determining the microion 
number density profiles, from which $\pi_e$ can be explicitly calculated.
In Sec.~\ref{results}, we demonstrate that, when implemented within Poisson-Boltzmann 
theory, Eq.~(\ref{pie3}) gives exactly the same results for $\pi_e$ as Eq.~(\ref{pie}), 
supporting the conjectured Eq.~(\ref{pie-def}).

\section{Computational Methods}\label{methods}

\subsection{Poisson-Boltzmann Theory}\label{PBtheory}

To evaluate our expression for the electrostatic component of the osmotic pressure 
of ionic microcapsules [Eq.~(\ref{pie3})], and thereby model swelling of these hollow
microgels, we computed microion number density profiles using Poisson-Boltzmann 
theory in the spherical cell model. Applications to interfaces between charged 
surfaces and ionic solutions~\cite{deserno2001cell,deserno2002,denton2010} have proven PB theory 
to be often a reasonable approximation for systems with monovalent (weakly correlated) 
microions~\cite{Ben_Yaakov_2009} and relatively low salt concentrations~\cite{denton2010,
Hallez}. Nevertheless, the mean-field approximation underlying the theory neglects 
correlations between microions, which can be important for multivalent microions. 

Combining the exact Poisson equation, 
\begin{equation}
\nabla^{2}\psi(r)= -4\pi\lambda_{B}[n_{+}(r)-n_{-}(r)-n_{f}(r)],
\label{psir}
\end{equation}
which relates the electrostatic potential to the total ion density,
with a mean-field Boltzmann approximation for equilibrium microion densities,
 \begin{equation}
n_{\pm}(r)=n_{0}\,\text{exp}[\mp\psi(r)],
\label{npm-Boltzmann}
\end{equation} 
yields the nonlinear PB equation (in three dimensions),
 \begin{equation}
\psi''(r)+\frac{2}{r} \psi'(r) = \kappa_0^2\sinh\psi(r) + 4\pi\lambda_B n_f(r),
\label{PB-eq}
\end{equation}
where $\kappa_0=\sqrt{8\pi\lambda_Bn_0}$ is the Debye screening constant in the 
salt reservoir and $n_f(r)$ is the fixed charge density of the microcapsule [Eq.~(\ref{nfr})].
Equation~(\ref{PB-eq}) can be numerically solved under the conditions of charge neutrality 
inside the spherical cell, $\psi'_{\rm out}(R)=0$, the symmetry requirement 
$\psi'_{\rm cav}(0)=0$, and the continuity of electrostatic potential at the 
inner and outer boundaries of the capsule shell, $\psi_{\rm cav}(a)=\psi_{\rm sh}(a)$
and $\psi_{\rm sh}(b)= \psi_{\rm out}(R)$, with the subscripts ``cav," ``sh," and 
``out" labelling the solutions in the cavity, shell, and outer regions, respectively.
In a closed cell with no salt, the PB equation [Eq.~(\ref{PB-eq})] becomes:
\begin{equation}
\psi''(r)+\frac{2}{r}\psi'(r) = -\kappa^2\ e^{-\psi(r)} + 4\pi\lambda_B n_f(r),
\label{PB-eq-closed} 
\end{equation}  
where $\kappa=\sqrt{4\pi\lambda_B Zn_m}$ is the screening constant in the 
cell~\cite{deserno2001cell}.
Although then determined by the electroneutrality condition ($Zn_m=n_+$), $\kappa$ 
is still physically associated with screening of macroions by counterions.
This association becomes evident when the PB equation is linearized around
the average potential in the cell, in which case $\kappa$ plays the role of the 
inverse Debye screening constant in the resultant Yukawa effective pair potential
between macroions~\cite{deserno2002,denton2010}.
Once the numerical solution of Eq.~(\ref{PB-eq}) or (\ref{PB-eq-closed}) 
is obtained, the electrostatic component of the osmotic pressure 
of a microcapsule follows directly from substituting $\la N_{\pm}\ra_{\rm cav}$, 
$\la N_{\pm}\ra_{\rm sh}$, and $\la r^2_{\pm}\ra_{\rm sh}$ into Eq.~(\ref{pie3}).

\subsection{Molecular Dynamics Simulations}

\begin{figure}[t!]
\includegraphics[width=0.9\columnwidth,angle=0]{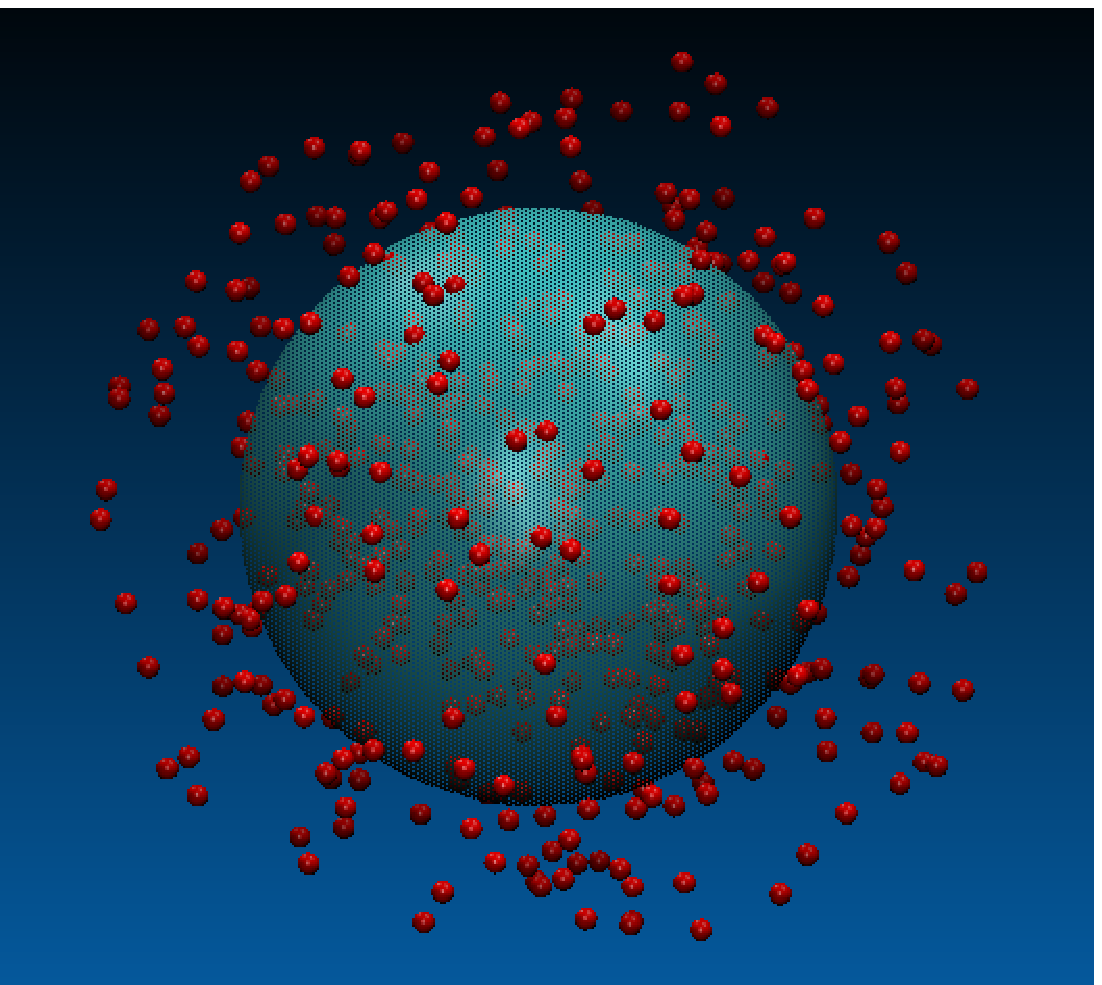}
\vspace*{-0.2cm}
\caption{Snapshot from molecular dynamics simulation of a penetrable ionic microcapsule 
(large cyan sphere) and mobile counterions (small red spheres) in a spherical cell.}
\label{fisnap}
\end{figure}

To validate our theoretical predictions for swelling of ionic microcapsules,
we performed molecular dynamics (MD) simulations in the cell model (Fig.~\ref{figmodel}),
using the LAMMPS package~\cite{plimpton1995,lammps}. In the constant-$NVT$ ensemble,
we simulated salt-free suspensions, containing only monovalent counterions.
Our method easily generalizes, however, to salty suspensions with multivalent microions.
We initialized the counterions on the sites of a face-centered cubic lattice in a cubic box 
with fixed boundary conditions and then allowed them to move according to Newton's equations 
of motion with mutual Coulomb pair forces and an ``external" electrostatic force 
generated by the charged microcapsule at the center of the cell. To model the influence 
of the latter force, we used the same technique as in ref.~\cite{Denton-Tang-jcp2016},
subjecting the counterions to a radial electric field, given by Eq.~(\ref{Efr}). 
To confine the counterions within the cell, we applied an additional short-range 
repulsive Lennard-Jones wall potential at the spherical boundary. Figure~\ref{fisnap} 
displays a snapshot from a typical simulation.

The equations of motion were integrated using the velocity-Verlet algorithm with 
a constant average temperature maintained via a Nos\'e-Hoover thermostat. After 
allowing the system to reach equilibrium, we accumulated a histogram of the 
counterion number density profile, from which we obtained the average numbers 
of counterions in the shell and cavity, $\la N_{+}\ra_{\rm sh}$ and 
$\la N_{+}\ra_{\rm cav}$, and the mean-square radial distance inside the shell, 
$\la r^{2}\ra_{\rm sh}$. Using these quantities, we computed the electrostatic component 
of the osmotic pressure $\pi_{e}$ of a microcapsule [Eq.~(\ref{pie3})]. 
We emphasize that all of the theoretical results and computational methods 
developed here for ionic microcapsules (hollow microgels) are equivalent to 
those developed for regular ionic microgels in the limit as the cavity vanishes.

\section{Results and Discussion}\label{results}

Swelling behavior of ionic microcapsules depends on many system parameters, including 
temperature, pH, ionic strength, concentration, and single-particle properties. 
Here we explore the dependence of the equilibrium swelling ratio on
microcapsule concentration, architecture, and charge at fixed temperature.
Following the approach established in refs.~\cite{Denton-Alziyadi2019} and
\cite{Denton-Tang-jcp2016}, we consider, for illustration, a uniformly charged 
spherical microcapsule comprising a total of 
$N_{\rm mon}=2\times 10^5$ monomers cross-linked into $N_{\rm ch}=100$ chains. 
In the dry (collapsed) state, we take the microcapsule inner and outer radii 
to be $a_{0}=10$ nm and $b_{0}=12.5$ nm, consistent with random close-packing 
of monomers of radius 0.15 nm.

To model aqueous solutions at $T=293$ K, we set the Bjerrum length $\lambda_{B}=0.714$ nm
and assume a Flory solvency parameter $\chi=0.5$. For simplicity, we consider 
salt-free solutions in which the only microions present are monovalent counterions.
As discussed in Sec.~\ref{suspension}, the point charge microion model is justified 
in this case, since for typical diameters of hydrated counterions~\cite{dzubiella2010} 
the electrostatic energy at contact is usually $> 2$ $k_BT$.
Elsewhere, we explored dependence of microcapsule swelling on salt 
concentration~\cite{Wypysek-Scotti2019}. For these system parameters, we aim first, 
to validate our approach by testing the predictions of the PB theory against 
MD simulations for the same model and second, to explore swelling trends 
with respect to variation of system parameters. To this end, we compute 
counterion density profiles, osmotic pressures, and equilibrium swelling ratios 
of ionic microcapsules modeled as described in Sec.~\ref{models}.

\begin{figure}[t!]
\includegraphics[width=1.0\columnwidth,angle=0]{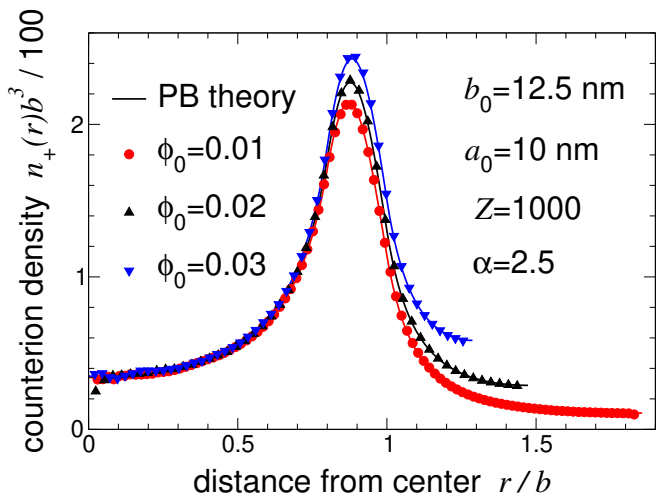}
\vspace*{-0.2cm}
\caption{Counterion number density profiles around an ionic microcapsule of dry 
inner and outer radii $a_{0}=10$ nm and $b_{0}=12.5$ nm, swollen radii $a=25$ nm 
and $b=31.25$ nm (swelling ratio $\alpha=2.5$), and valence $Z=1000$ 
in an aqueous solution at $T=293$ K ($\lambda_B=0.714$ nm) from MD simulations 
(symbols) and PB theory (curves) in the cell model at dry volume fractions 
$\phi_0=0.01$, 0.02, and 0.03.
}\label{fignr}
\end{figure}

For a given set of system parameters, the equilibrium microion number density profiles 
$n_{\pm}(r)$ satisfy Eqs.~(\ref{npm-Boltzmann}) and (\ref{PB-eq}), or (\ref{PB-eq-closed})
in the case of no salt. We solved these relations numerically in Mathematica using 
the {\tt NDSolve} and {\tt FindRoot} methods. Figure~\ref{fignr} shows the resulting 
counterion density profiles $n_+(r)$, together with our simulation data, for dry 
microcapsule volume fractions $\phi_{0}=(b_{0}/R)^3=0.01$, 0.02, and 0.03 and a 
swelling ratio $\alpha=2.5$, corresponding to swollen volume fractions 
$\phi=(b/R)^3=\alpha^3\phi_0 = 0.15625$, 0.3125, and 0.46875. The close agreement 
between predictions of the PB theory and our MD simulation data validates our methods. 
Note that the counterion density is relatively flat near the center and edge 
of the cell, where the electric field vanishes, and peaks inside the shell, 
where the oppositely charged fixed charge of the microcapsule is localized.

\begin{figure}[t!]
\includegraphics[width=1.0\columnwidth,angle=0]{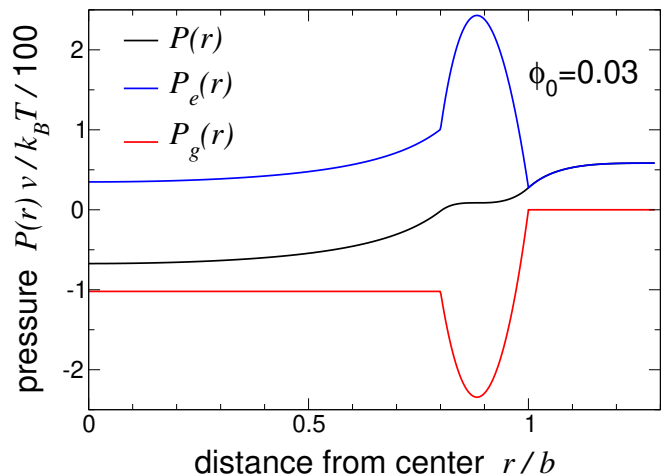}
\vspace*{-0.2cm}
\caption{Radial profiles of pressure around an ionic microcapsule from PB theory 
in the cell model for same microcapsule parameters as Fig.~\ref{fignr}
and dry volume fraction $\phi_0=0.03$. Plotted are the total pressure $P(r)$
(black), the electrostatic component $P_e(r)$ (blue), and the gel component $P_g(r)$ (red).
}\label{figptotr}
\end{figure}
  
\begin{figure}[t!]
\includegraphics[width=1.0\columnwidth,angle=0]{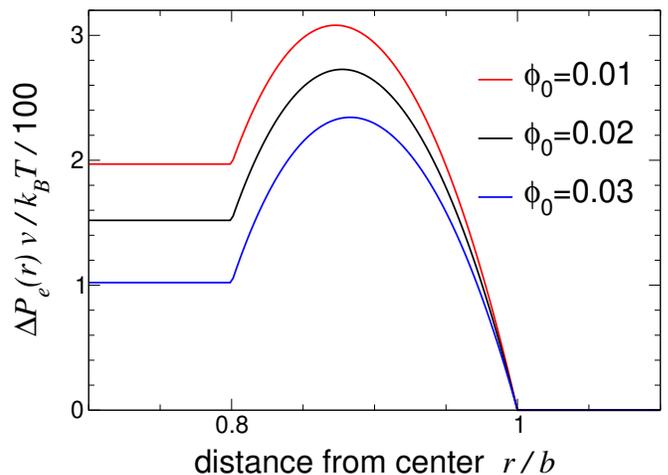}
\vspace*{-0.2cm}
\caption{Radial profiles of deviation of electrostatic component of pressure 
$\Delta P_e(r)=-P_g(r)$ around an ionic microcapsule from PB theory in the cell model 
[Eqs.~(\ref{Pgr2}) and (\ref{Delta-Per1})] for same microcapsule parameters 
as Fig.~\ref{fignr} and dry volume fractions $\phi_0=0.01$, 0.02, and 0.03.
}\label{figper}
\end{figure}

We also computed radial profiles of the total pressure $P(r)$ and its electrostatic 
and gel components, $P_e(r)$ and $P_g(r)$, by evaluating Eqs.~(\ref{Ptot}), (\ref{Per}), 
and (\ref{Pgr2}), using the electrostatic potential $\psi(r)$ predicted by PB theory 
in the cell model, obtained from numerically solving Eqs.~(\ref{npm-Boltzmann}) and 
(\ref{PB-eq-closed}).
Figure~\ref{figptotr} displays our numerical results for the same microcapsule 
parameters as in Fig.~\ref{fignr} and at dry volume fraction $\phi_0=0.03$. 
The total pressure (black curve in Fig.~\ref{figptotr}) varies continuously 
across the microcapsule shell, decreasing monotonically from a positive value 
at the cell boundary ($r=R$) to a negative value at the cell center ($r=0$). Note that 
$P(r)<0$ is physically sensible, since the condition for stability is that the 
divergence of the pressure tensor vanish [Eq.~(\ref{divergence-sphere})]. Interestingly,
$P(r)$ is relatively level near the boundary and center of the cell and within the shell, 
but varies most rapidly near the edges of the shell.

The gel component of the pressure (red curve in Fig.~\ref{figptotr}) is strictly 
negative inside the microcapsule and constant inside the cavity, which is devoid 
of gel. Note that the magnitude of $P_g(r)$ in the shell is strongly correlated 
with the densities of fixed charge and counterions. Correspondingly, the 
electrostatic component of the pressure (blue curve in Fig.~\ref{figptotr}), 
which is the difference between the total pressure and its gel component, is 
everywhere positive for these parameters, varying rapidly within the charged shell 
and relatively gradually within the cavity, which harbors some fraction of the counterions. 

Figure~\ref{figper} illustrates the variation with dry volume fraction of the deviation
of the electrostatic component from the total pressure $\Delta P_e(r)$, which equals 
the negative of the gel component. Interestingly, this deviation, which peaks near the 
center of the shell, decreases in overall magnitude with increasing dry volume fraction. 
This trend can be understood by considering the variation of the {\it local} Debye
screening length, defined as $\lambda_D(r)\equiv 1/\sqrt{4\pi\lambda_B n_+(r)}$. 
With increasing $\phi_0$, the counterion density $n_+(r)$ inside the shell increases,
thus shortening $\lambda_D$ and enhancing electrostatic screening of the fixed charge.
As a result, the electrostatic component of the pressure is reduced. This trend is
qualitatively consistent with a similar trend observed in ref.~\cite{Wypysek-Scotti2019} 
with increasing salt concentration. Note that the different values of $\Delta P_e(r)$
at the inner and outer surfaces of the shell ($r=a$ and $r=b$) result from
different local Donnan equilibria at these two interfaces.

\begin{figure}[t!]
\includegraphics[width=1.0\columnwidth,angle=0]{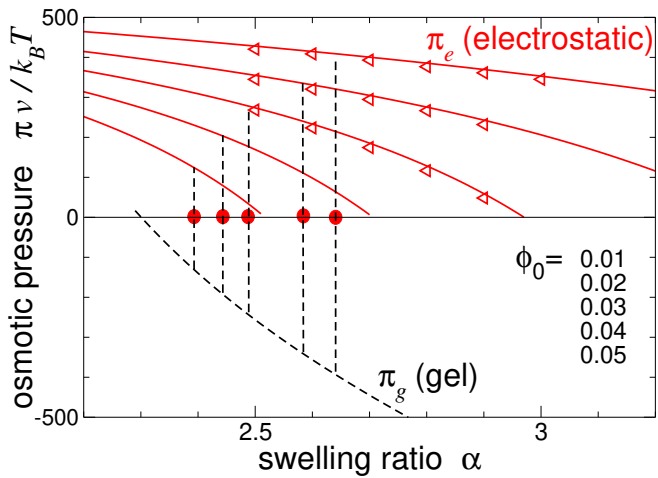}
\vspace*{-0.2cm}
\caption{Components of osmotic pressure vs.~swelling ratio $\alpha$ 
of a single ionic microcapsule in the cell model at dry volume fractions 
$\phi_0=0.01-0.05$ (top to bottom). Electrostatic component $\pi_e$ from 
Poisson-Boltzmann theory [Eq.~(\ref{pie3})] (solid red curves) and from 
MD simulations (open symbols). Gel component $\pi_g$ (dashed black curve) 
from Flory-Rehner theory [Eq.~(\ref{pig2})]. At equilibrium swelling, 
the total osmotic pressure vanishes: $\pi_e+\pi_g=0$ (filled symbols).
Dashed vertical lines connect corresponding points on the $\pi_e$ and $\pi_g$ curves.
System parameters are same as in Fig.~\ref{fignr}.}
\label{figpi}
\end{figure}
 
To compute the electrostatic component of the total osmotic pressure $\pi_e$, 
we evaluated Eq.~(\ref{pie3}), obtaining $\la N_+\ra_{\rm cav}$, $\la N_+\ra_{\rm sh}$,
and $\la r^2_+\ra_{\rm sh}$ by integrating the equilibrium counterion number 
density profile $n_+(r)$ predicted by PB theory in the cell model. As a consistency 
check, we independently determined $\pi_e$ from our MD simulations, which naturally 
include interparticle correlations, by computing histograms for the requisite 
ensemble averaged properties from the counterion density profile. As shown in Fig.~\ref{figpi},
over a range of swelling ratios, theory (solid red curves) and simulation (open symbols) 
agree nearly exactly for $\pi_e$, further validating our methods and also confirming
the accuracy of the mean-field Boltzmann approximation underlying PB theory, 
implying relatively weak correlations among the monovalent counterions. 
Remarkably, averaging the radial component of the electrostatic pressure tensor $\Delta P_e(r)$ 
over the microcapsule volume [Eqs.~(\ref{Pgr2})-(\ref{pie})] yields numerically exactly 
the same result as the thermodynamic electrostatic osmotic pressure $\pi_e$ computed 
from Eq.~(\ref{pie3}), consistent with our earlier findings for regular 
microgels~\cite{Denton-Alziyadi2019}.

Figure~\ref{figpi} further demonstrates that the electrostatic component of the 
osmotic pressure monotonically decreases with increasing swelling ratio $\alpha$,
for a given dry volume fraction $\phi_0$, and also decreases with increasing 
$\phi_0$ for a given $\alpha$. The first trend can be attributed to the declining 
contribution to $\pi_e$ from the self-energy [first term on right side of 
Eq.~(\ref{pie3})] as the density of fixed charge in the shell decreases upon swelling. 
The second trend results from an increasing fraction of counterions inside the 
microcapsule with increasing $\phi_0$ and the fact that interior counterions
act to neutralize the fixed charge. 

Figure~\ref{figpi} also shows the gel component of the osmotic pressure 
(dashed black curve), as predicted by Flory-Rehner theory [Eq.~(\ref{pig2})].
With increasing swelling ratio, $\pi_g$ monotonically decreases. For uncharged 
microcapsules with these parameters, the equilibrium swelling ratio 
(where $\pi_g=0$) is $\alpha_{\rm eq}\simeq 2.3$. For ionic microcapsules, the 
electrostatic component of the osmotic pressure drives additional swelling. 
The solid red symbols in Fig.~\ref{figpi} identify the swelling ratios 
at which the total osmotic pressure vanishes: 
$\pi_{e}(\alpha_{\rm eq})+\pi_{g}(\alpha_{\rm eq})=0$. 

\begin{figure}[t!]
\includegraphics[width=1.0\columnwidth,angle=0]{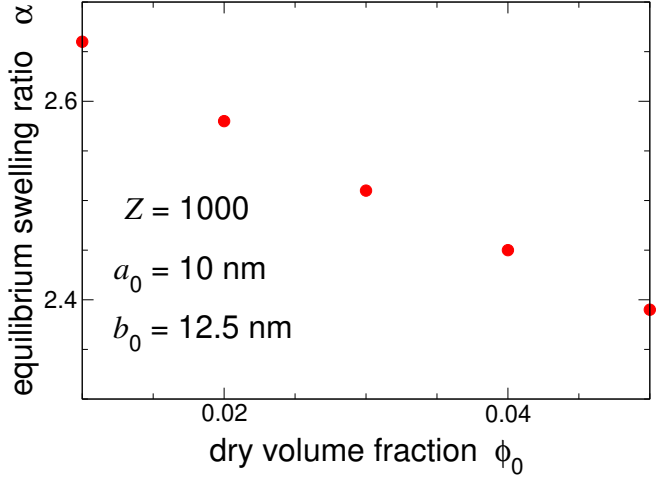}
\vspace*{-0.2cm}
\caption{Equilibrium swelling ratio $\alpha$ of ionic microcapsules vs.~dry 
volume fraction $\phi_0$ for same system parameters as in Fig.~\ref{fignr}. 
With increasing $\phi_0$, microcapsules steadily deswell.}
\label{alpha_vs_phi0}
\end{figure}

These data are replotted in Fig.~\ref{alpha_vs_phi0} to illustrate the 
variation of equilibrium swelling ratio with dry volume fraction.
As $\phi_0$ increases from 0.01 to 0.05, the equilibrium value of $\alpha$ 
steadily decreases from 2.66 to 2.39, while the actual volume fraction, 
$\phi=\phi_0\alpha^3$, increases from about 0.19 to 0.68. 
Thus, we find that deswelling of ionic microcapsules upon crowding 
occurs at concentrations well below those at which steric interactions 
between particles are significant. We attribute such deswelling, therefore, 
to crowding-induced redistribution of counterions, consistent with conclusions
from previous studies of ionic microgels with no cavities~\cite{Denton-Tang-jcp2016,weyer-denton2018}.

\begin{figure}[t!]
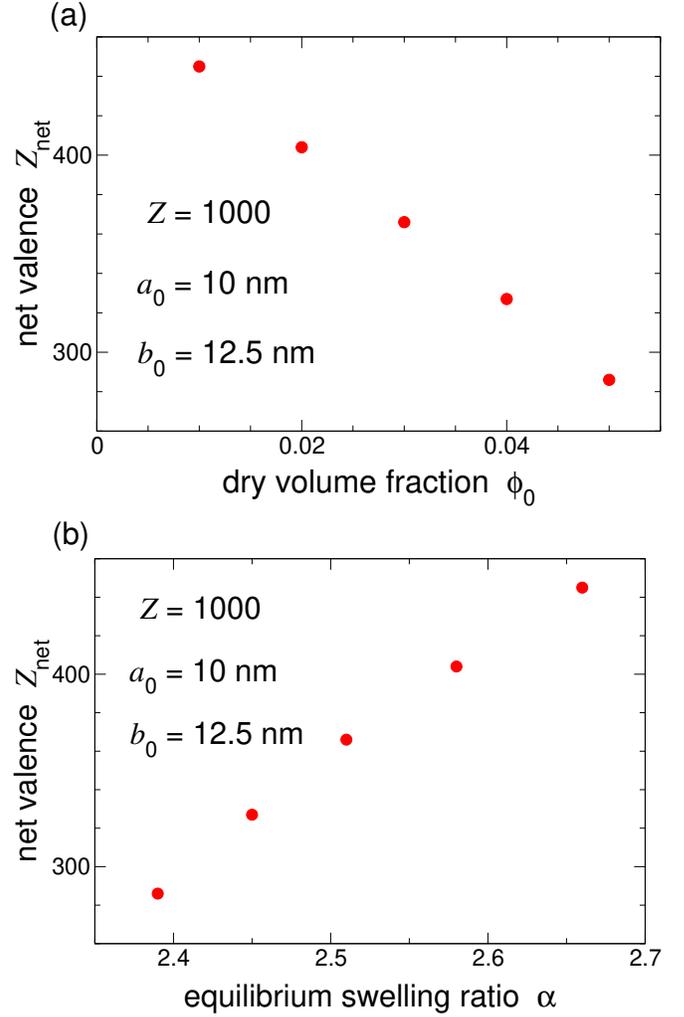

\includegraphics[width=1.0\columnwidth,angle=0]{znet_vs_dry_volume_fraction_mc_1000.eps}
\\[1ex]
\includegraphics[width=1.0\columnwidth,angle=0]{znet_vs_alpha_mc_1000.eps}
\vspace*{-0.2cm}
\caption{Dependence of microcapsule mean net valence $Z_{\rm net}$ on 
(a) dry volume fraction $\phi_0$ and (b) equilibrium swelling ratio $\alpha$.
Other system parameters are same as in Fig.~\ref{fignr}.}
\label{Znet}
\end{figure}

To quantify our interpetation of why $\pi_e$ and $\alpha$ decrease with increasing $\phi_0$, 
we computed the mean net valence of a microcapsule, $Z_{\rm net}\equiv Z-\la N_+\ra_{\rm in}$,
where $\la N_+\ra_{\rm in}$ is the mean number of interior counterions, i.e.,
in the shell and cavity regions ($r\le b$). 
Figure~\ref{Znet}(a) shows the variation of $Z_{\rm net}$ (at equilibrium swelling) 
with $\phi_0$, confirming that the mean number of interior counterions increases 
with increasing concentration. Figure~\ref{Znet}(b) illustrates the corresponding 
increase of $Z_{\rm net}$ with increasing equilibrium swelling ratio. We conclude 
that the counterion distribution strongly influences equilibrium swelling 
of ionic microcapsules, even at relatively low concentrations.

\begin{figure}[t!]
\includegraphics[width=1.0\columnwidth,angle=0]{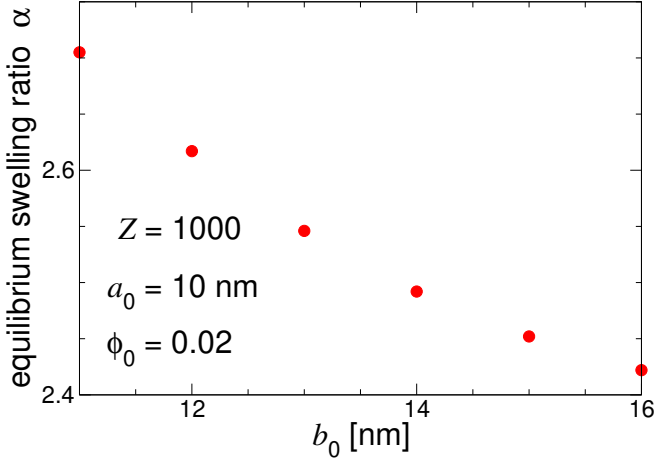}
\vspace*{-0.2cm}
\caption{Dependence of equilibrium swelling ratio $\alpha$ on dry outer radius $b_0$ 
of ionic microcapsules with fixed dry inner radius $a_0=10$ nm, valence $Z=1000$, 
and dry volume fraction $\phi_0=0.02$.
Other system parameters are same as in Fig.~\ref{fignr}.}
\label{alpha_vs_b0}
\end{figure}
\begin{figure}[t!]
\includegraphics[width=1.0\columnwidth,angle=0]{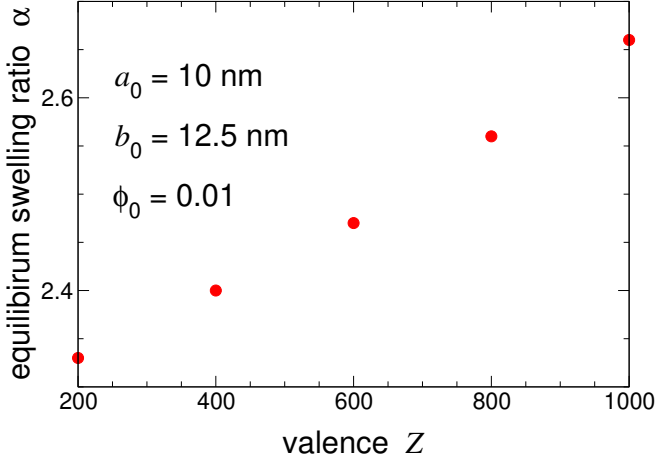}
\vspace*{-0.2cm}
\caption{Dependence of equilibrium swelling ratio $\alpha$ on valence $Z$
of ionic microcapsules with fixed dry inner radius $a_0=10$ nm, dry outer
radius $b_0=12.5$ nm, and dry volume fraction $\phi_0=0.01$.
Other system parameters are same as in Fig.~\ref{fignr}.
With increasing $Z$, microcapsules steadily swell.}
\label{alpha_vs_Z}
\end{figure}

To analyze the influence of shell thickness on swelling of ionic microcapsules, 
we computed the equilibrium swelling ratio as a function of outer dry radius for 
a given fixed inner dry radius. As seen in Fig.~\ref{alpha_vs_b0}, at fixed dry
volume fraction, the equilibrium swelling ratio monotonically decreases with 
increasing shell thickness. From the statistical mechanical expression for $\pi_e$ 
[Eq.~(\ref{pie3})], this trend clearly results from a decrease in the electrostatic 
component of osmotic pressure as the fixed charge distribution becomes more diffuse.

Finally, Fig.~\ref{alpha_vs_Z} illustrates the dependence of microcapsule 
swelling on valence, which in experiments can be varied by changing 
the solution pH. With increasing $Z$, at fixed concentration ($\phi_0$),
the equilibrium value of $\alpha$ monotonically increases. This trend can be 
explained by inspection of Eq.~(\ref{pie3}), which reveals that the 
electrostatic component of osmotic pressure $\pi_e$ increases with increasing $Z$, 
amplifying electrostatics and favoring swelling.

\section{Conclusions}\label{conclusions} 

In summary, we developed and applied theoretical and computational methods for modeling 
osmotic pressure and equilibrium swelling behavior of ionic microcapsules in suspension.
For a simple physical model of an ionic microcapsule confined to a spherical cell,
with fixed charge uniformly distributed throughout the hydrogel shell, 
we analyzed the pressure tensor and derived exact statistical mechanical relations 
for the electrostatic component of the osmotic pressure of a single particle.
Using Poisson-Boltzmann theory and molecular dynamics simulation, we computed 
radial profiles of the counterion density and pressure in the spherical cell model. 
Close agreement between predictions from the two methods validates our approach
and provides a consistency check.

Combining the electrostatic component with the gel component of the osmotic pressure
of a microcapsule, which we modeled via the Flory-Rehner theory of polymer networks, 
we computed the equilibrium swelling ratio of microcapsules and investigated the
dependence on system parameters, including particle concentration, shell thickness, 
and valence. For an illustrative set of parameters representative of concentrated 
suspensions of highly charged microcapsules with swollen radii in the range of 
30-35 nm and relatively thin shells (few nm in thickness), our analysis demonstrates 
that the counterion density sharply peaks inside the shell, but levels off near 
the center and edge of the cell, where the electric field tends to zero.
 
Our calculations of counterion density and equilibrium swelling further show 
that a redistribution of counterions between the inside and outside of the shell
causes microcapsules of a given valence to deswell with increasing concentration 
and with increasing shell thickness, while particles of a given concentration 
swell with increasing valence. Our results complement those from an earlier 
study~\cite{Wypysek-Scotti2019}, which predicted the influence of ionic strength 
on swelling in qualitative agreement with experiment. 

The trends identified here suggest further possibilities for tuning swollen sizes 
of microcapsules by changing environmental properties, such as concentration or pH. 
Thus, our modeling approach can help to guide future synthesis and characterization 
of ionic microcapsules, which may have practical applications, e.g., to biosensing 
and drug delivery. Finally, our methods can be easily extended to other soft 
colloidal particles with different architectures, such as 
cylindrical microgels~\cite{mohammed-alan-cylindrical}.

\vspace*{0.2cm}
\acknowledgments
We thank Andrea Scotti for helpful discussions.
MOA thanks Shaqra University for financial support. ARD acknowledges support of
the National Science Foundation (Grant No.~DMR-1928073). This work used resources 
of the Center for Computationally Assisted Science and Technology (CCAST)
at North Dakota State University, made possible in part by NSF MRI Award No.~2019077.

\begin{center}
{\bf DATA AVAILABILITY}
\end{center}

The data that support the findings of this study are available from the corresponding author
upon reasonable request.



%

\end{document}